\documentclass[aps,prd,notitlepage,nofootinbib,superscriptaddress,twocolumn]{revtex4-1}
\usepackage[utf8]{inputenc}
\usepackage{amsmath}
\usepackage{amssymb}
\usepackage{amsfonts}
\usepackage{newtxtext,newtxmath}
\usepackage{bm}
\usepackage{graphicx}

\usepackage[usenames,dvipsnames]{xcolor}
\usepackage{color}
\usepackage[colorlinks=true,linkcolor=blue,urlcolor=blue,citecolor=blue]{hyperref}

\usepackage{slashed}
\usepackage[english]{babel}
\usepackage{dcolumn}
\usepackage{blindtext}
\usepackage{epsfig}
\usepackage{pifont}
\usepackage{dsfont,mathrsfs}
\usepackage{cancel}
\usepackage{bigints}
\usepackage{accents}
\usepackage{soul}
\usepackage{multirow}
\usepackage{ulem}
\usepackage{simpler-wick}
\usepackage{float}

\def\nn{\nonumber}
\def\beq{\begin{equation}}
\def\eeq{\end{equation}}
\def\beqa{\begin{eqnarray}}
\def\eeqa{\end{eqnarray}}
\def\ban{\begin{eqnarray*}}
\def\ean{\end{eqnarray*}}
\def\bi{\begin{itemize}}
\def\ei{\end{itemize}}
\begin{document}

\title{Causality violation and the speed of sound of hot and dense quark 
matter in the Nambu--Jona-Lasinio model}

\author{Arthur E.B. Pasqualotto}\email{ arthur.pasqualotto@acad.ufsm.br}
\affiliation{Departamento de F\'{\i}sica, Universidade Federal de Santa
  Maria,  97105-900 Santa Maria, RS, Brazil}
  
\author{ Ricardo L. S. Farias}\email{ ricardo.farias@ufsm.br}
\affiliation{Departamento de F\'{\i}sica, Universidade Federal de Santa
  Maria, 97105-900 Santa Maria, RS, Brazil}
  
\author{William R. Tavares} \email{william.tavares@posgrad.ufsc.br}
\affiliation{Departamento de F\'{\i}sica, Universidade Federal de Santa
  Catarina, 88040-900 Florian\'{o}polis, SC, Brazil}
\affiliation{Departamento de Física Teórica, Universidade do Estado do Rio de Janeiro, 20550-013 Rio de Janeiro, RJ, Brazil}        
  
\author{Sidney S. Avancini}\email{sidney.avancini@ufsc.br}
\affiliation{Departamento de F\'{\i}sica, Universidade Federal de Santa
  Catarina, 88040-900 Florian\'{o}polis, SC, Brazil}  
  
\author{ Gast\~ao Krein}\email{gastao.krein@unesp.br}
\affiliation{Instituto de F\'isica Te\'orica, Universidade Estadual Paulista, Rua Dr. 
Bento Teobaldo Ferraz, 271 - Bloco II, 01140-070 S\~ao Paulo, S\~ao Paulo, Brazil}
  
\begin{abstract} 
The Nambu--Jona-Lasinio model is widely used to study strong-interaction phenomena in vacuum and quark matter. Since the model is nonrenormalizable, one needs to work 
within a specific regularization scheme to obtain finite results. Here we show that a 
commonly used cutoff regularization scheme leads to unphysical results, such as 
superluminal speed of sound and wrong high-temperature behavior of the specific 
heat and other thermodynamical quantities. Such a troublesome 
feature of the cutoff regularization invalidates the model for temperature 
and baryon density values relevant to the phenomenology of heavy-ion collisions and 
compact stars. We show that the source of the problems stems from cutting off momentum 
modes in finite integrals depending on thermal distribution functions in the grand 
canonical potential. The problems go away when taking into account the full momentum 
range of those integrals. Explicit examples are worked out in the SU(2)-flavor version 
of the model.
\end{abstract}

\pacs{}
 
\maketitle

%
\section{Introduction}

Signals of a phase change, from a cold and low-density hadronic 
phase to a hot and low-density quark-gluon phase, has been identified in 
several observables collected from heavy-ion collision experiments at the 
Relativistic Heavy Ion Collider (RHIC) and the Large Hadron Collider; Refs.~\cite{Snellings:2011sz,Busza:2018rrf,Elfner:2022iae}
provide reviews with extensive lists of references. A basic quantity characterizing 
new phases of matter produced in heavy-ion collisions is the equation of state 
(EOS)~\cite{Braun-Munzinger:2015hba}. From the EOS, one can compute the speed of 
sound $c_s$, which can be used to identify phase changes in the matter produced 
in those collisions~\cite{Sorensen:2021zme}. The speed of sound is also relevant 
in the study of the cold and dense matter found in compact stars; $c_s$ controls 
the stiffness of the EOS~\cite{Li:2019fqe,Alford:2019oge,LopeOter:2019pcq,Li:2021sxb,
Greif:2018njt,Duarte:2020kvi,Kojo:2020krb,Stone:2019blq,Tews:2018iwm,Kojo:2014rca}, 
a property that constrains the mass-radius relation of such stars~\cite{Alford:2013aca,
Bedaque:2014sqa,VanOeveren:2017xkv,Margaritis:2019hfq,Otto:2019zjy,Landry:2020vaw,
Jokela:2020piw,Moustakidis:2016sab,Jimenez:2021wil,Annala:2019puf,Hippert:2021gfs,
Ferreira:2020kvu,Tan:2021ahl,Tews:2019cap,Jaikumar:2021jbw,Constantinou:2021hba,
Han:2019bub,Ma:2018qkg,McLerran:2018hbz}.

Quantum chromodynamics (QCD) is the fundamental theory governing 
strongly interacting matter. A first-principles QCD computation of the EOS at finite
temperatures has been performed with lattice QCD simulations using Monte Carlo 
methods. But when dealing with matter at baryon densities relevant to compact stars,
neutron stars in particular, sign problems obstruct the use of those methods 
and alternative theoretical methods are required. At low baryon densities~$n_B$, densities 
much smaller than the 
nuclear saturation density $n_0=0.16~\text{fm}^{-3}$, it is possible
to construct an EOS with chiral perturbation theory (ChPT)~\cite{Jimenez:2021wil,Tews:2012fj,Hebeler:2013nza} 
or through many-body nuclear models and constraints
from experimental data~\cite{Kojo:2020krb}. In the very-high-density regime,
e.g., $n_B\gtrsim 40 n_0$, 
the QCD asymptotic freedom property enables 
the use of weak-coupling perturbation theory~\cite{Kojo:2020krb,Freedman:1976dm,Kurkela:2009gj,
Fraga:2013qra,Tews:2018kmu}. Computations using this approach~\cite{Kurkela:2009gj,Tews:2018kmu} found a value for the
speed of sound below the conformal limit $c_s^2 = 1/3$, as expected for low-density matter.
In the intermediate-density regime, relevant for the phenomenology of neutron stars, interpolation and extrapolation 
strategies have been used; for a recent review, see Ref. \cite{Baym:2017whm}. 
For such densities, there is an ongoing debate~\cite{Bedaque:2014sqa,Moustakidis:2016sab,Reed:2019ezm,
Kanakis-Pegios:2020jnf,Drischler:2020fvz} regarding the upper bound of $c^2_s$.

The speed of sound has also been computed at finite temperature and
chemical potential in several contexts using different phenomenological approaches. 
This is the case of heavy-ion collisions, where the entropy per baryon, $s/n_B$, is approximately 
constant in the whole expansion stage of the collision~\cite{Huovinen:2014woa}. 
Estimations of entropy per baryon have been obtained within the range
$s/n_B=17.5$ and $331.6$ for collisions with center-of-mass energy in the range between 
$\sqrt{s_{NN}}=7.7$ and $200 \text{ GeV}$ \cite{Motta:2020cbr}, 
with data provided by the STAR Collaboration in Refs.~\cite{STAR:2008med,STAR:2017sal}. 
Another example is the envisaged experiments at the Nuclotron-based
Ion Collider fAcility, at which it is 
expected to obtain entropy to baryon density ratios in the range $s/n_B = 4-11$ 
with the collision energies in the range 
$\sqrt{s_{NN}} = 4-11~\text{ GeV}$~\cite{Blaschke:2017lzo}.
In the context of Taylor expansion coefficients in lattice QCD, 
it is possible to construct an EOS with values of $s/n_B$ in the range  
$s/n_B \sim 30-300$  for the conditions achieved in experiments at the 
RHIC, SPS, and AGS~\cite{Ejiri:2005uv,Huovinen:2014woa,Bluhm:2007nu}. Lower values
of $s/n_B$ can be obtained looking at the different stages of formation 
of a proto-neutron star described in terms of isentropic lines~\cite{Steiner:2000bi,Mariani:2016pcx,Mariani:2017fqr,
Constantinou:2014hha,Constantinou:2015mna}. In the first stage, 
where $s/n_B \sim 1$, there is an abundance of trapped neutrinos, followed by the process of
deleptonization, where the star reaches its maximum inner temperature with $s/n_B \sim 2$. 
In the final stage, with the tendency of the  diffusion of neutrinos and cooling, the star 
can reach $s/n_B\sim 0$.

Alternatively to the high temperature and/or chemical potential systems, there are other 
environments that are explored in the context of heavy-ion 
collisions. Because of the nontrivial structure of QCD vacuum at high temperatures, 
classical solutions called sphalerons take place~\cite{RevModPhys.53.43,PhysRevLett.60.1902}
and can induce, through the Adler-Bell-Jackiw anomaly, 
an imbalance between right-handed and left-handed quarks.  Such configurations can be present in 
$C$ and $CP$ \cite{Fukushima:2008xe,Kharzeev:2009pj} violating processes. 
Such a situation can happen in the quark-gluon plasma when a strong magnetic field 
perpendicular to the reaction plane is produced in a peripheral heavy-ion collision~\cite{Kharzeev:2004ey}, 
 generating a vector current in the same direction. This effect is known as the chiral magnetic 
effect~\cite{Fukushima:2008xe}, and has been extensively explored theoretically and 
experimentally~\cite{Kharzeev:2015znc,Huang:2015oca,Zhao:2019hta,STAR:2021mii,yamanaka2022unobservability,Yamanaka:2022vdt}.
Some of these aspects encouraged the investigation of the effects in the effective
quark masses and chiral density in the context of parallel electromagnetic
fields~\cite{Ruggieri:2016xww,Ruggieri:2016lrn}. Neutron stars, in principle, could
be affected by the chiral imbalance since they are subject to high 
densities and/or strong magnetic fields~\cite{sym12122095}. Furthermore, lattice
QCD simulations of a chiral-imbalanced medium, with the imbalance set by a
chiral chemical potential~$\mu_5$, do not suffer from the sign problem~\cite{Yamamoto:2011gk},
a welcome feature for comparisons of effective models~\cite{Farias:2016let}. 

These different high-density scenarios impose technical difficulties in their
modeling with a nonrenormalizable model, as for example with the SU(2) Nambu--Jona-Lasinio (NJL) 
model~\cite{Nambu:1961tp,Nambu:1961fr,Klevansky:1992qe}  and its extensions to more 
flavors~\cite{Hatsuda:1994pi,BUBALLA_2005}. The NJL model has been widely used to study 
several aspects related to the QCD chiral transition, in vacuum, in matter at finite
temperature and density~\cite{Klevansky:1992qe,Hatsuda:1994pi,BUBALLA_2005} and external 
magnetic field~\cite{Bandyopadhyay:2020zte}.  Since the model is nonrenormalizable, 
the regularization scheme becomes part of the model, in that the regulator cannot be removed. 
However, careless use of a regularization scheme can lead to gross violations of basic physical 
principles. This happens with one of the most used regulation schemes, namely, cutoff 
regularization, or traditional regularization scheme (TRS)~\cite{Farias:2016let},  
in which divergent integrals are regulated with a momentum cutoff~$\Lambda$. 
In the present paper, we show that this regularization scheme, as used in many publications, e.g.~\cite{Wang:2021mfj,Wu:2020qgr,Zhao:2019xqy,Sheng:2020hge,Sheng:2021evj,doi:10.7566/JPSCP.20.011009,Yu:2015hym,Wu:2020knd,Yang:2019lyn,Cui:2017ilj},
can lead to causality violation through a superluminal speed of sound of quark matter and
can also lead to a conformality violation and wrong behavior of the specific heat and other thermodynamic
quantities at high temperatures. We show that the source of such troublesome features is 
related to the use of a cutoff in finite integrals in the grand canonical 
potential. A known example is the Stefan-Boltzmann limit of thermodynamical 
quantities. One obtains the Stefan-Boltzmann limit only when the finite thermal 
contributions depending on the Fermi-Dirac distributions are integrated over 
the full momentum range~\cite{Costa:2009ae,Costa:2010zw, Marty:2013jga}; the 
Stefan-Boltzmann limit is not obtained when those contributions are integrated with 
the same ultraviolet cutoff $\Lambda$ used to regulate divergent integrals~\cite{Zhuang:1994dw}. 

Similar problems are normally ignored when considering finite-density and 
chiral-imbalanced systems. For chiral-imbalanced systems modeled within the NJL model
with a chiral chemical potential $\mu_5$, there exist just a few works concerning the role 
of regularization prescriptions. The importance of an adequate regularization resides in 
the fact that the vacuum contribution of the thermodynamical potential in the NJL model 
turns out to be entangled with a contribution depending on the chiral chemical potential 
$\mu_5$ in a divergent contribution. One can get finite results adopting the TRS 
scheme~\cite{Yu:2015hym}. However, such a scheme~\cite{Yu:2015hym,Chao:2013qpa,Yu:2014sla} 
leads to results in disagreement with the lattice QCD simulations of Refs.~\cite{Braguta:2015owi,
Braguta:2015zta}; this is also true when using the TRS in PNJL models~\cite{Ruggieri:2011xc,
Fukushima:2010fe,Cui:2016zqp}. It is worth mentioning that the linear sigma model coupled 
to quarks~\cite{Chernodub:2011fr,Ruggieri:2011xc}, which is a renormalizable model, also leads
to results disagreeing with those lattice results. One can get agreement with the lattice
results when one uses the regularization named medium separation scheme~\cite{Farias:2016let},
in that one separates the vacuum contributions from the medium. Besides all of these issues, 
the chiral chemical potential can also induce divergences in the chiral charge density, due to 
the nonconservation of the chiral current in the physical limit, 
which imposes a normalization procedure \cite{sym12122095,Ruggieri:2020qtq}. 
There are some other applications studying
the chiral symmetry restoration and the QCD phase diagram with different approaches and models in 
Refs.~\cite{Braguta:2016aov,Lu:2016uwy,Yang:2019lyn,Shi:2020uyb}.

In this work, guided by all this phenomenology, we study within an SU(2) NJL model
the effects of regulating or not with of an ultraviolet cutoff finite integrals contributing to the 
thermodynamic potential of quark matter under two scenarios: finite quark chemical 
potential and the chiral chemical potential. The isentropic lines are obtained in both 
$T-\mu_q$ and $T-\mu_5$ planes and the speed of sound is calculated. To this end, we present in Sec.~II 
the formalism of the SU(2) NJL model at finite temperature and chemical potential. 
We present the explicit expressions for the thermodynamical quantities depending on $T$ and $\mu_q$ 
in Sec.~II. A, and those depending on $T$ and $\mu_5$ in Sec.~II. B. 
We present numerical results in Sec.~III. In Sec.~IV we present our conclusions.

%
\section{NJL model at finite temperature and density}

We use a NJL model defined by the following Lagrangian 
density~\cite{Nambu:1961tp,Nambu:1961fr,Klevansky:1992qe,Hatsuda:1994pi,BUBALLA_2005}:
\begin{align}
 \mathcal{L}_{\rm NJL}=&\bar{\psi}\left(i\slashed{\partial}-\hat{m} \right)\psi 
 + G\left[(\bar{\psi}\psi)^2+(\bar{\psi}i\gamma_5{\bm\tau}\psi)^2\right],
 \label{NJL-lag}
\end{align}
\noindent 
where $\psi = (u \; d)^{T}$ represents the $u$ (up) and $d$ (down) quark isospin doublet, 
$\hat{m} = \text{diag}(m_u \; m_d)$ is the quark mass matrix, $G$ the coupling strength, 
and ${\bm\tau} = (\tau^1, \tau^2, \tau^3)$ are the isospin Pauli matrices. 

Like with most quantum field theoretical models, ultraviolet divergences occur in the process 
of calculating physical quantities within the NJL model. In order to give physical sense to the 
results derived from the model, one needs a regularization procedure. Since the NJL~model is 
nonrenormalizable, the regularization procedure becomes an integral part of the model. In this 
work, we employ a three-dimensional momentum cutoff scheme, by far the most frequently
used scheme in the NJL model. We refer to it as the traditional regularization 
scheme~(TRS), a previously defined nomenclature~\cite{Farias:2016let}.

In the following, we present the derivation of the expressions for the thermodynamical quantities 
within the~TRS. We~consider first the expressions for quark matter at finite-temperature~$T$
and quark baryon chemical potential~$\mu_q$. Next, we derive the expressions for quark matter 
at finite~$T$ and a chiral imbalance of right- and left-handed quarks by fixing the 
imbalance with a quark chiral chemical potential~$\mu_5$. The chemical potentials 
enter in the grand canonical potential~$\Omega$, from which one derives the thermodynamical 
quantities. In~the~imaginary time formalism of finite temperature 
field theory, one incorporates $\mu_q$~and~$\mu_5$ by adding  to the Euclidean action 
corresponding to the Lagrangian density ${\mathcal L}_{\rm NJL}$ the following 
terms:
\begin{align}
S_{\mu_q}   & = \int^\beta_0\!d\tau \int\!d^3x \; \bar{\psi} 
\mu_q \gamma_0\psi, \\
S_{\mu_5} & = \int^\beta_0\!d\tau \int\!d^3x \; \bar{\psi} \mu_5 \gamma_0 \gamma_5 \psi,
\end{align}
with the quark fields obeying antiperiodic boundary conditions in $\beta = 1/T$, 
namely, $\psi({\bm{x}}, 0) = - \psi({\bm{x}}, \beta)$. We compute~$\Omega$ in the 
mean-field approximation and work in the isospin symmetry limit $m_u = m_d = m_0$. 

The thermodynamical quantities of interest in this paper are the speed of sound~$c_s$, 
specific heat~$c_V$, interaction strength (or trace anomaly)~$\Delta$, 
and conformal measure~$C$. They are defined in terms of partial derivatives of quantities defining 
the EOS, the pressure~$P = - \, \Omega$, the energy density~$\epsilon$,
the entropy density~$s$ and quark number density~$\rho$: 
\begin{align}
P_N &= P(T,\mu) - P(0,0), \label{PN} \\[0.25true cm]
\epsilon &= - P_N + T s + \mu \rho,
\label{epsilon} \\[0.25true cm]
s &= -\left(\frac{\partial \Omega}{\partial T}\right)_{\mu},
\hspace{1.0cm}
\rho = - \left(\frac{\partial \Omega}{\partial T}\right)_{\mu}, 
\label{def-s-rho}
\end{align}
where $M = M(T,\mu)$ is the constituent quark mass, the solution of the gap equation. 
Here, $\mu$ stands for either $\mu_q$ or $\mu_5$ and thereby $\rho$ stands for $\rho_q$ 
or~$\rho_5$, respectively. From these, one can then readily compute $c_s$, $c_V$, 
$\Delta$, and~$C$ as follows: 
\begin{align}  
c_s^2 &= - \left(\frac{\partial P_N}{\partial \epsilon}\right)_{s/\rho} 
= \frac{\rho^2\Upsilon_{T,T}-2(s\rho)\Upsilon_{T,\mu}-s^2\Upsilon_{\mu,\mu}}
{(\epsilon+P_N)(\Upsilon_{T,T}\Upsilon_{\mu,\mu}-\Upsilon_{\mu,T}^2)}, 
\label{cs2} \\
c_V &= \left(\frac{\partial \epsilon}{\partial T}\right)_{\rho} 
= T\left(\frac{\partial s}{\partial T}\right) 
= T \left(\Upsilon_{T,T}-\frac{\Upsilon_{\mu,T}^2}{\Upsilon_{\mu,\mu}}\right), 
\label{cV} \\
\Delta &= \epsilon - 3P_N, \quad C = \frac{\Delta}{\epsilon},
\label{D-C}
\end{align}
where $\Upsilon_{\alpha,\beta}=\partial^2 P_N/\partial \alpha \partial\beta$ as defined 
in Refs.~\cite{Marty:2013ita,Floerchinger:2015efa}. 

Another quantity of interest in this study is the quark condensate for a given flavor:
\beq 
\langle \bar\psi_f \psi_f\rangle_{T,\mu} = - N_c 
\int \frac{d^4k}{(2\pi)^4} \, {\rm Tr}_{\rm D}\, S^{(f)}_{T,\mu}(k), 
\eeq 
where $S^{(f)}_{T,\mu}(k)$ is a $T$- and $\mu$-dependent 
flavor$-f$ quark propagator in momentum space, 
${\rm Tr}_{\rm D}$ means trace over Dirac indices and 
$N_c = 3$ is the number of colors. We recall that the 
quark condensate is the (quasi)order parameter of the QCD 
chiral phase transition; in the chiral limit, it is a true 
order parameter.

%
\subsection{Finite quark chemical potential~\texorpdfstring{$\mu_q$}{mu}}
\label{subsec:mu}

For a nonzero quark chemical potential~$\mu_q$, the mean-field thermodynamic 
potential corresponding to the Lagrangian in Eq.~\eqref{NJL-lag} can be written as
\begin{align} 
\Omega(T,\mu_q) &= \frac{(M-m_0)}{4G} - 2 N_c N_f \int_{\Lambda}\frac{d^3k}{(2\pi)^3} \,  
\omega(k) \nn \\
&\;\;\;\, - 4 N_c N_f  T \int_{\Lambda_T} \frac{d^3k}{(2\pi)^3} \,  
\Bigl[ \log\left(1+e^{-(\omega(k)+\mu_q)/T}\right) \nn \\  
&\;\;\;\,  + \log\left(1+e^{-(\omega(k)-\mu_q)/T}\right) \Bigr],
 \label{omega-muq}
\end{align}
where $\omega(k) = (k^2+M^2)^{1/2}$ and $N_f = 2$ are,  respectively, the 
color and flavor numbers. To~assess the consequences of cutting off momentum modes in finite integrals, 
we introduced in the above $\Lambda_T$ to indicate either $\Lambda_T = \Lambda$ or 
$\Lambda_T \rightarrow \infty$. In the former case, one is cutting off momentum modes in finite 
integrals that do not require a regularization, as the explicit dependence of the 
integrals on $T$ and~$\mu_q$ makes them finite. In the latter case, all momentum 
modes contribute to the finite integrals. This procedure is similar to the one used 
in Ref.~\cite{Yu:2015hym}, although in that reference the authors restricted their
analyses to the quark condensate as a function of~$\mu_5$ only and have not 
addressed causality violation issues, the main topic of the present study.  

The gap equation for the constituent quark mass~$M$ is 
given by
\begin{align}
M &= m_0 + 4 N_c N_f G \int_{\Lambda}\frac{d^3k}{(2\pi)^3}\, \frac{M}{\omega(k)} 
\nn \\ 
&\;\;\;\, - \, 4 N_c N_f G \int_{\Lambda_T} \frac{d^3k}{(2\pi)^3} \, \frac{M}{\omega(k)}
\left[( n_-(k)  + n_+(k) \right],\label{gap_mu}
\end{align}
where $n_\mp(k)$ are the quark and antiquark Fermi-Dirac distributions:
\beq
n_\mp(k) = \frac{1}{e^{(\omega(k) \mp \mu_q)/T} + 1  }. 
\label{FD-dist}
\eeq 
While the first term in the gap equation \eqref{gap_mu} is ultraviolet divergent,
the second is finite due to the Fermi-Dirac distributions that provide a natural cutoff 
for the high momentum modes. Even for $T=0$ the second term is finite, since 
$n_-(T=0,\mu_q) = \theta(k_F - k)$ and $n_+(T=0,\mu_q) = 0$, where 
$k_F = (\mu^2_q - M^2)^{1/2}$ is the Fermi momentum.

To compute the thermodynamical quantities of interest via Eqs.~(\ref{cs2})--(\ref{D-C}), 
with $\mu = \mu_q$ and $\rho = \rho_q$, we need the expressions for the entropy
density~$s$ and quark number density~$\rho_q$: 
\begin{align}
\hspace{-0.25cm}s &= 2N_cN_f \! \int_{\Lambda_T} \! \frac{d^3k}{(2\pi)^3} 
\Biggl\{ 
\log \bigl[ 1 - n_-(k) \bigr] \bigl[ 1 - n_+(k) \bigr]
\nonumber\\
&\;\;\;\, - \, \frac{\omega(k)}{T} \bigl[ n_-(k) + n_+(k) \bigr] 
+ \frac{\mu_q}{T} \bigl[ n_-(k) - n_+(k) \bigr] 
\Biggr\}, 
\label{s-muq}
\\[0.2true cm]
\hspace{-0.25cm}\rho_q &= - 2 N_c N_f \! \int_{\Lambda_T} \! \frac{d^3k}{(2\pi)^3}
\bigl[  n_-(k) - n_+(k)  \bigr].
\label{rho-q}
\end{align}
Here, we have only finite integrals as  they involve 
the Fermi-Dirac distributions $n_{\mp}(k)$, which depend 
on $T$ and $\mu$.

%
\subsection{Finite quark chiral chemical potential \texorpdfstring{$\mu_5$}{mu5} }
\label{subsec:mu5}

The mean-field grand canonical potential at finite quark chiral
potential~$\mu_5$ can be written as:
\begin{align}
&\Omega(T,\mu_5) = \frac{(M-m_0)}{4G} -N_cN_f\sum_{s=\pm 1}
\int_{\Lambda}\frac{d^3k}{(2\pi)^3} \, \omega_s(k) \nn \\ 
&\hspace{0.75cm} \, - \, 2 N_c N_f T \sum_{s=\pm 1}
\int_{\Lambda_T} \frac{d^3k}{(2\pi)^3} \, \log\left(1 + e^{- \omega_s(k)/T} \right),
\label{omega-mu5}
\end{align}
where $\omega_s(k)=((k + s \mu_5)^2 + M^2)^{1/2}$ are the eigenvalues of the Dirac
Hamiltonian for quark helicities~$s = \pm 1$. Again, one has a divergent integral 
and a finite integral, in which we introduced $\Lambda$ and $\Lambda_T$. The gap 
equation for the constituent quark mass~$M$, in this case, is given by: 
\begin{align} 
M &= m_0 + 2 N_c N_f G \sum_{s=\pm 1}
\int_{\Lambda}\frac{d^3k}{(2\pi)^3}\frac{M}{\omega_s(k)} \nn \\
&\;\;\;\, - 4 N_c N_f G \int_{\Lambda_T}\frac{d^3k}{(2\pi)^3} \, 
\frac{M}{\omega_s(k)} \, n_s(k),
\end{align}
where $n_s(k)$ is the Fermi-Dirac distribution
\begin{equation} n_s(k) = \frac{1}{e^{ \omega_s(k)/T} + 1}. 
\end{equation}

The entropy density is given by
\begin{align}
s &= 2 N_c N_f \sum_{s=\pm 1} \int_{\Lambda_T} \frac{d^3k}{(2\pi)^3} 
\, \left[ 
\frac{\omega_s(k)}{T} \, n_s(k) \right. \nn \\
&\;\;\;\, \left. + \, \log \left( 1+e^{- \omega_s(k)/T}\right) \right].
\label{s-mu5}
\end{align} 
The fact that we used $s$ to indicate both the entropy density and quark chirality
should not cause confusion as their meaning should be clear by the context in
which they appear in the equations. Now, contrary to the baryon charge, the chiral 
charge is not a conserved quantity for massive quarks; thereby, the chiral
density~$\rho_5$ is not finite in the chirally 
broken phase of the NJL model 
and one needs a prescription to give physical meaning to~$\rho_5$. Here we follow 
the prescription of Ref.~\cite{Ruggieri:2020qtq} and redefine $\rho_{5}$ as 
$\rho_{5N} = - \partial \Omega_N/\partial \mu_5$, where $\Omega_N(T,\mu_5) =
\Omega(T,\mu_5) - \Omega_N(0,\mu_5)$, so that
\begin{align}
\rho_{5N} &= N_c N_f \sum_{s=\pm 1}\int_{\Lambda_T} \frac{d^3k}{(2\pi)^3}
\, \frac{s(k+s\mu_5)}{\omega_s(k)} 
\Bigl( 1 - 2 \, n_s(k) \Bigr) .
\end{align}
For a thorough discussion on the meaning of such a renormalization in QCD, 
we direct the reader to Ref.~\cite{Ruggieri:2020qtq}. The expressions for the 
speed of sound, specific heat, trace anomaly, and conformal measure are given
by Eqs.~(\ref{cs2})-(\ref{D-C}) with $\mu$ replaced by $\mu_5$ 
and $\rho$ by $\rho_{5N}$.

%
\section{Numerical Results}

We adopted the following values for the free parameters of the model: $\Lambda = 587.9$ MeV, 
$m_0=5.6$ MeV and $G=2.44/\Lambda^2$. These values were chosen to reproduce the phenomenological 
vacuum values of the pion decay constant $f_{\pi}=92.4$~MeV, pion mass $m_{\pi}=135$~MeV,  
and the chiral condensate~$\langle \bar{u}{u}\rangle=(-240.8 \, \text{MeV})^{1/3}$~\cite{BUBALLA_2005}. 
In the following, we present the results for the speed of sound ~$c_s$, for zero chemical potentials $\mu_q = 0$ 
and $\mu_5 = 0$. Then, in the next two subsections, we consider nonzero values of $\mu_q$ and 
$\mu_5$ and present results for $c_V$, $\Delta$ and~$C$. 

\begin{figure}[htb]   
\centering
\includegraphics[width=8.5cm]{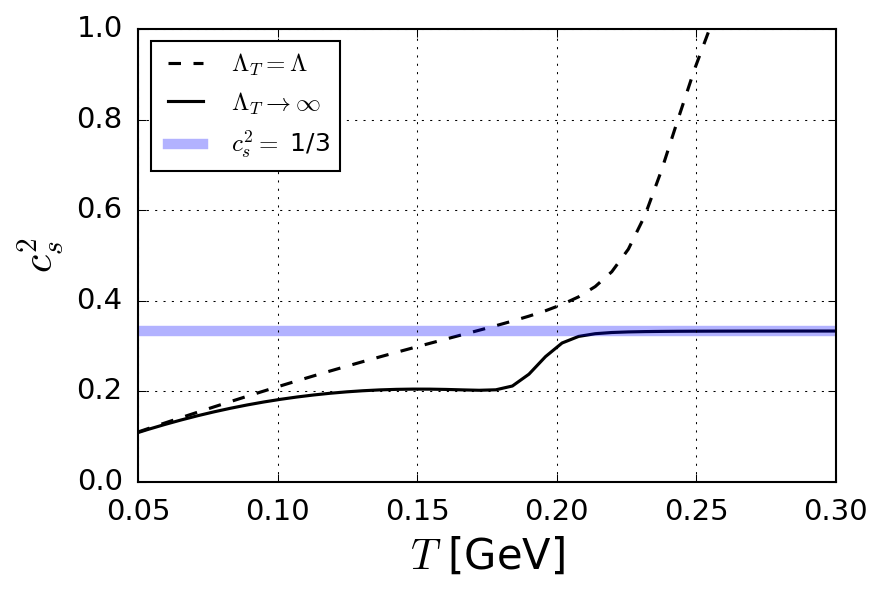} 
\caption{ The speed of sound squared as a function of the temperature~$T$ 
for zero chemical potentials, $\mu_q~=~\mu_5~=~0 $. 
}
\label{fig:mu=0_cs2}
\end{figure} 

Figure~\ref{fig:mu=0_cs2} displays the temperature dependence of the 
squared speed of sound. The message the figure conveys is clear: Using $\Lambda_T = \Lambda$,
with which one cuts off not only divergent integrals but also finite ones, the speed of sound 
exceeds the speed of light, thereby violating causality. Causality violation occurs for 
temperatures $T \gtrsim 255 \text{ MeV}$, temperatures well in the range of those achieved 
in a heavy-ion collision. Moreover, the conformal 
limit $c_s^2=1/3$ is violated for temperatures $T > 170$~MeV. 
Now, when one keeps the whole contribution of
the quark and antiquark Fermi-Dirac distributions, i.e. when using $\Lambda_T \to \infty$, 
the speed of sound  approaches~$c_s^2 = 1/3$, the conformal value,  
which is the correct value of the speed of sound of a high-temperature 
relativistic free gas of fermions--the  
Stefan-Boltzmann (SB) limit. 

The relationship between the cutoff and the 
wrong behavior of the speed of sound can be qualitatively 
understood as follows. The system cannot build enough pressure to
reach the SB limit when one eliminates quark high-momentum modes; 
for $T \gtrsim 150$~MeV, instead of increasing toward the SB limit, 
the pressure decreases with~$T$. The energy density also decreases 
with~$T$ for $T \gtrsim 150$~MeV, but it decreases more rapidly than 
the pressure, which explains the rapid rise of $c^2_s$ for 
those temperatures. As we show in the following 
sections, where we consider the effects of finite $\mu_q$ and
$\mu_5$, we obtain similar unphysical results not only for 
the speed of sound but also for other thermodynamical 
quantities, the reason for that is essentially the same 
as here for $\mu_q = \mu_5 =0$.

\subsection{ Causality violation on the \texorpdfstring{$T - \mu_q $}{Tmuq} plane }

\begin{figure}[h!]
 \centering
 \includegraphics[width=0.45\textwidth]{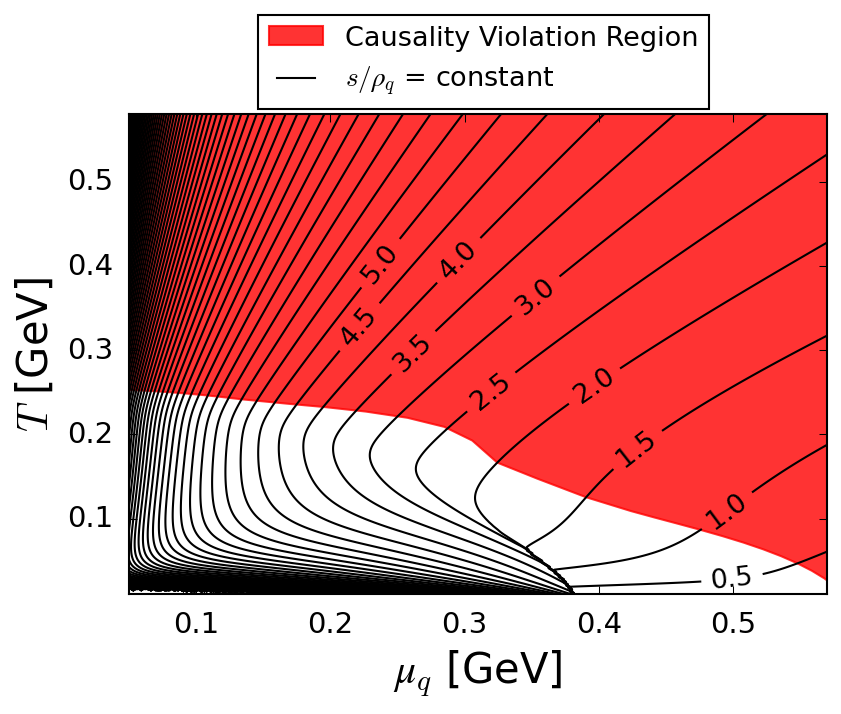}
 \caption{
The red-shaded area identifies the region of causality 
 violation ($c^2_s > 1$) in the $T - \mu_q$ plane for  
 constant values of entropy to quark baryon density ratios 
 (black curves). These results are obtained using a three-dimensional cutoff in 
 finite integrals, i.e.  $\Lambda_T = \Lambda$.
 }
 \label{fig:T_mu_violation_curve}
\end{figure}

We consider the $T$ and $\mu_q$ dependence of $c^2_s$ using 
a three-dimensional momentum cutoff in both the divergent and finite integrals, 
i.e. we use $\Lambda_T = \Lambda$. Figure~\ref{fig:T_mu_violation_curve} displays 
the results for different values of the entropy to quark baryon density ratio, $s/\rho_q$. 
The red-shaded area represents the regions for which $c^2_s > 1$. It is evident 
that $c^2_s$ is nonphysical for a wide range of phenomenologically relevant values
of $T$ and $\mu_q$. Within this regularization scheme, the model can be safely used in the 
phenomenology of relativistic heavy-ions collisions ($\mu_q \sim 0$) only for temperatures up to 
$T \sim 250$ MeV, whereas in the phenomenology of quark or hybrid stars ($T \sim 0$) it can be 
safely used only for quark chemical potentials up to $\mu_q \sim 500$~MeV (that correspond to the baryon density $\rho_B \approx 7 \rho_0$, where $\rho_0 = 0.16 fm^{-3}$ is the saturation density of nuclear matter). 

\begin{figure}[h!]
 \centering
 \includegraphics[width=0.45\textwidth]{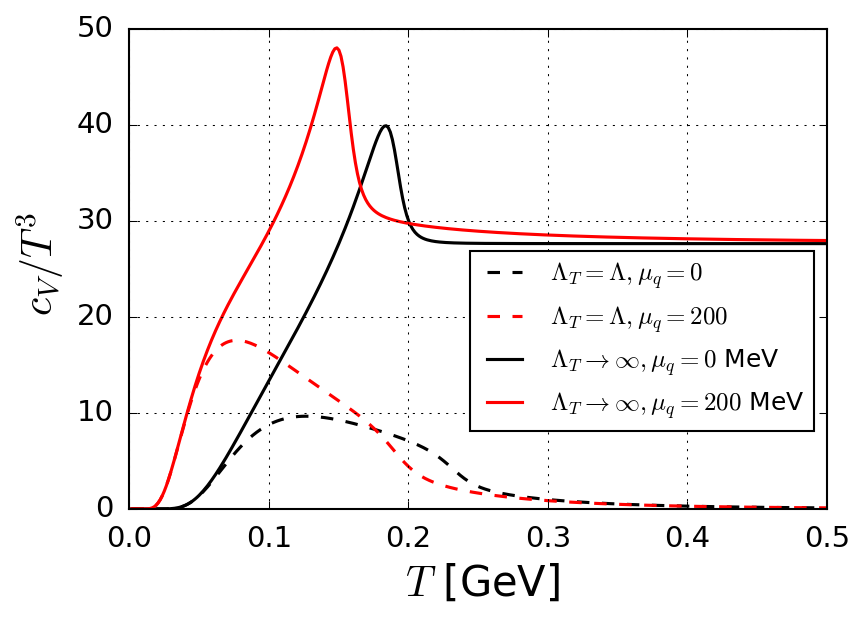}
 \caption{
 Specific heat normalized by $T^3$ as function 
 of the temperature for two values of the quark baryon chemical 
 potential $\mu_q$. 
 }
 \label{cv_mu}
\end{figure}

\begin{figure}[h!]
 \centering
 \includegraphics[width=0.45\textwidth]{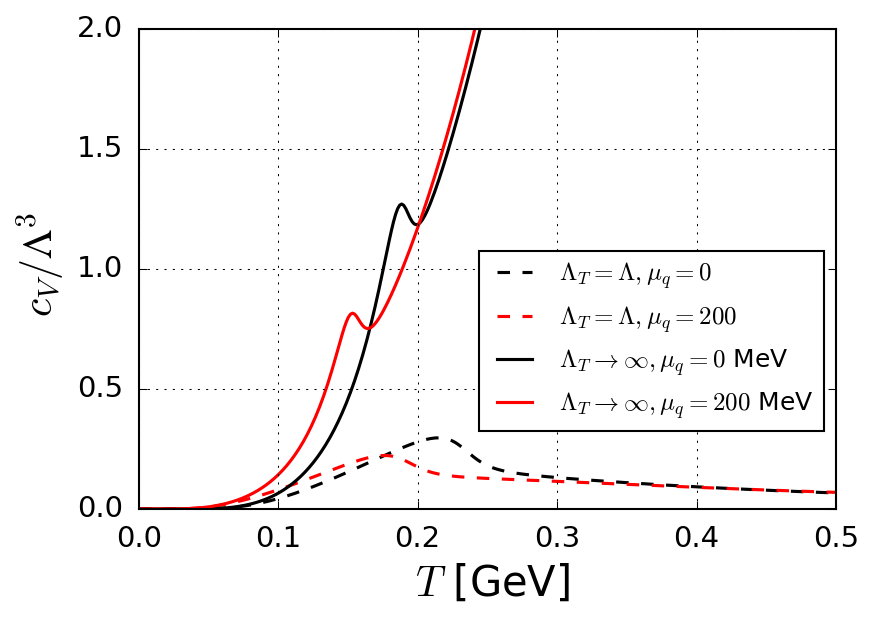}
 \caption{ 
 Specific heat normalized by $\Lambda^3$ as a function of the temperature for 
 two values of the quark baryon chemical 
 potential $\mu_q$. 
}
 \label{cv_mu2}
\end{figure}

Next, we consider the effect of the cutoff in the other
observables of interest by comparing results obtained with $\Lambda_T = \Lambda$ 
and $\Lambda_T \rightarrow \infty$. First, we consider the specific heat $c_V$ as
a function of the temperature for $\mu_q=0$ and $\mu_q = 200\text{ MeV}$. Figures~\ref{cv_mu} 
and \ref{cv_mu2} display, respectively, $c_V/T^3$ and $c_V/\Lambda^3$; we consider these 
two different normalizations to assess the high-temperature behavior of~$c_V$ better.
Using $\Lambda_T = \Lambda$, one obtains nonphysical temperature dependence for $c_V$ 
for temperatures $T \gtrsim 50 \text{ MeV}$ for both values of quark chemical 
potential considered. The fact that in this case  $c_V/T^3$ 
is bigger at low temperatures for $\mu_q=200$ MeV than for 
$\mu_q=0$
is mainly due to the normalization by $T^3$. In fact, 
$c_V$ increases smoothly at low $T$ as a function of $\mu_q$, but it increases 
in a very artificial way when compared to the case with $\Lambda_T\rightarrow\infty$, 
as we can see in Fig. \ref{cv_mu2}. The conclusion is that when using
$\Lambda_T=\Lambda$ one obtains a wrong high-temperature behavior of the specific heat.

\begin{figure}[h!]
 \centering
 \includegraphics[width=0.45\textwidth]{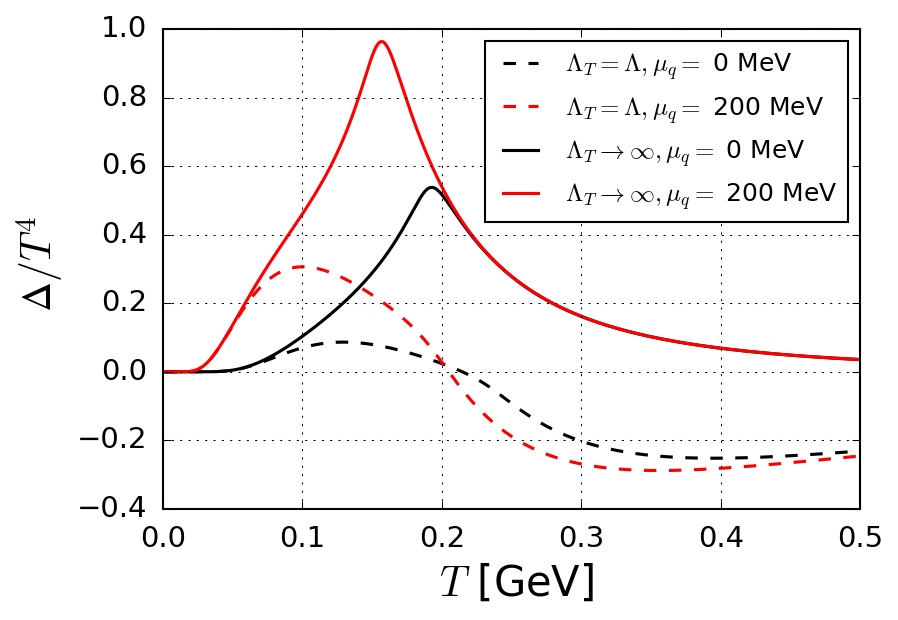}
 \caption{
 The normalized trace anomaly (or interaction measure) $\Delta$ 
 as a function of the temperature for two values of the 
 quark baryon chemical potential. 
 }
 \label{fig:T_mu_Delta}
\end{figure}

\begin{figure}[h!]
 \centering
 \includegraphics[width=0.45\textwidth]{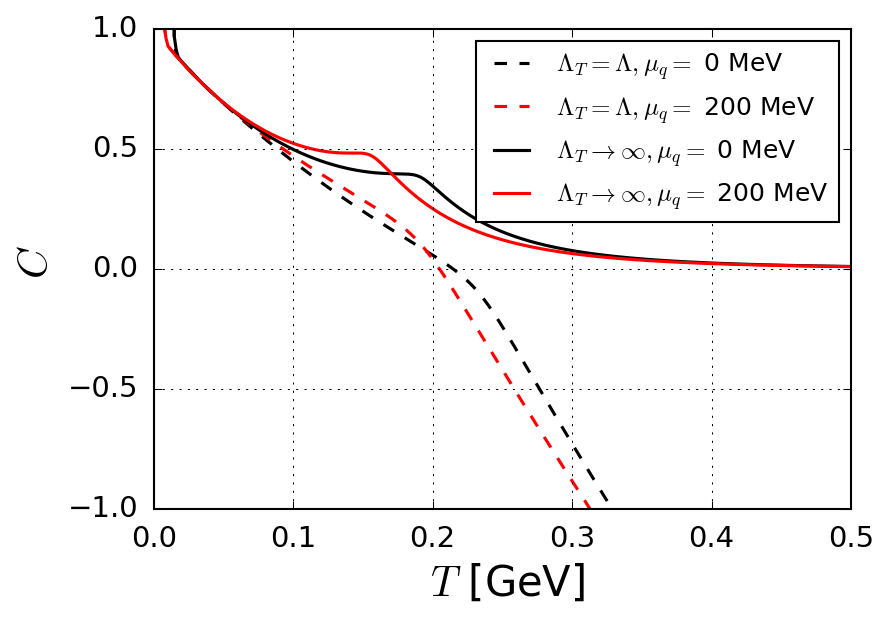}
 \caption{
 The conformal measure $C = \Delta/\epsilon$ 
 as a function of the temperature for two values of the quark chemical potential. 
 }
 \label{fig:T_mu_Delta2}
\end{figure}

Finally,  Figs.~\ref{fig:T_mu_Delta} and \ref{fig:T_mu_Delta2} display, respectively,  
the trace anomaly $\Delta$ normalized by $T^4$ and the conformal measure $C=\Delta/\epsilon$ as a function of temperature.
Both quantities are expected to tend asymptotically to zero at large $T$. Recall
that in the mean-field approximation, the effect of the interaction disappears at
high temperatures due to the chiral restoration; see the first terms in
Eqs.~(\ref{omega-muq}) and (\ref{omega-mu5}). Such a behavior happens
when we use $\Lambda_T \to \infty$, but it 
does not happen when one uses $\Lambda_T = \Lambda$, 
for both finite and zero quark chemical potential. The pattern found 
in the case of $\Lambda_T \to \infty$, in both quantities, indicates that 
for asymptotically high temperatures, the theory behaves as a free theory, 
as expected within this mean-field approximation.    
The results obtained with $\Lambda_T=\Lambda$ 
and $\Lambda_T\rightarrow\infty$ start to deviate from each other 
at temperatures close to $T = 100$~MeV, 
in both $\Delta$ and $C$. Moreover, for temperatures close to $T = 200$~MeV, both 
quantities become negative when using $\Lambda_T=\Lambda$. 
Recent works~\cite{Fujimoto:2022ohj,Pinto:2022lkv,Pinto:2022elo} show 
that both the trace anomaly and conformal measure are positive quantities at 
finite density, indicating that using $\Lambda_T = \Lambda$ leads to wrong 
high-temperature behavior for these quantities.

\begin{figure}[h!]   
 \centering
 \includegraphics[width=0.45\textwidth]{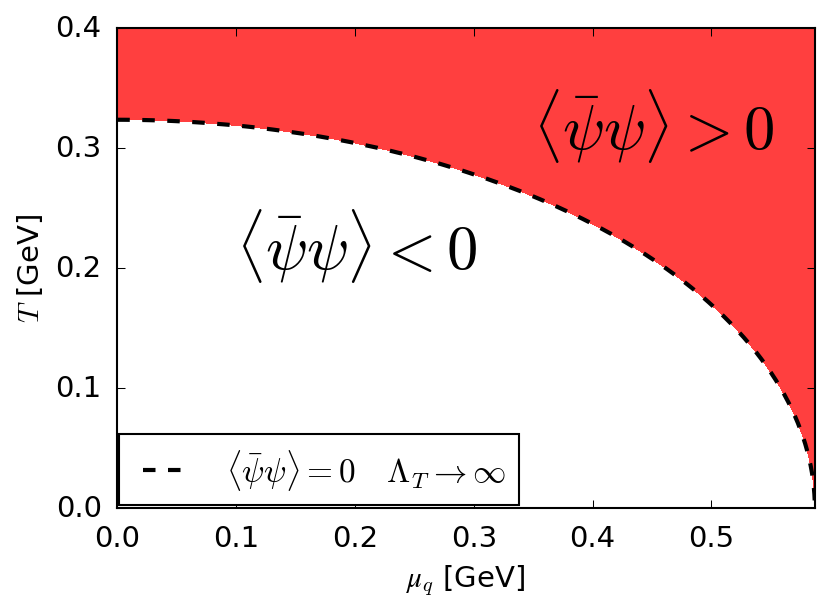} 
 \caption{ The quark condensate as a function of $T$ and $\mu_q$ computed with 
 $\Lambda_T \rightarrow  \infty$. }
 \label{fig:condensatemu}
\end{figure}

We close this section by showing in Fig.~\ref{fig:condensatemu} the quark condensate 
$\langle \bar\psi\psi\rangle$ computed with $\Lambda_T \rightarrow \infty$. The figure reveals that 
at both extremes of $T$ and $\mu_q$ relevant to the phenomenologies of heavy collisions 
and neutron starts, one obtains a condensate that is still negative (the white region). 
This means that the constituent quark  mass~$M > m_0$ at those extremes; when 
$\langle \bar\psi\psi\rangle > 0$, one can have $M < m_0$. 

\subsection{ Causality violation on the \texorpdfstring{$T- \mu_5$}{Tmu5} plane }

\begin{figure}[h!]   
 \centering
 \includegraphics[width=0.45\textwidth]{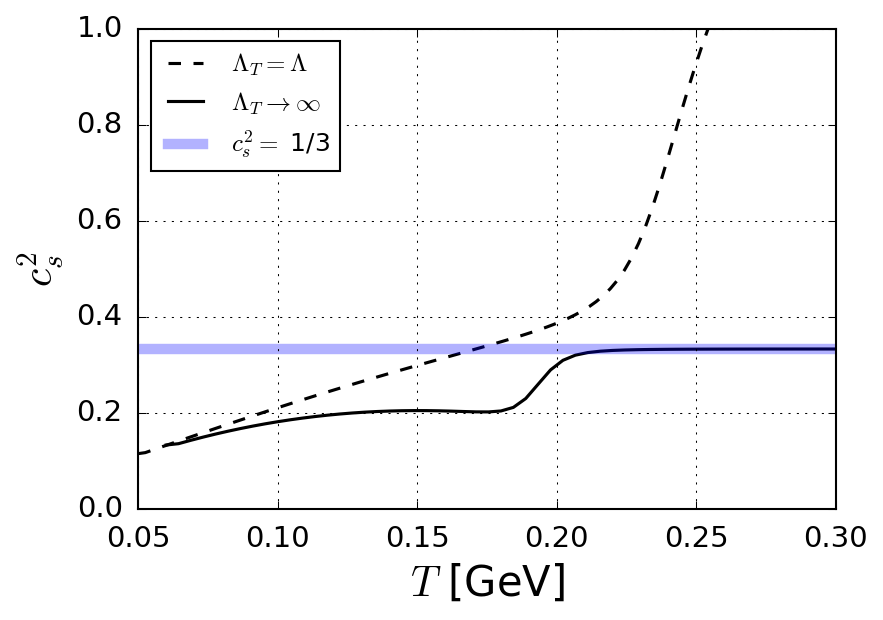}
 \caption{Speed of sound squared as a function of $T$ for $s/\rho_5 = 100 $. 
}
 \label{fig:mu=0_cvrho5}
\end{figure}

In this section, we discuss causality violation for 
the case of finite chiral chemical potential, $\mu_5$. Again, we start analyzing the effect of using $\Lambda_T = \Lambda$ in the
expression for the speed of sound; the results are displayed in Fig.~\ref{fig:mu=0_cvrho5} 
for $s/\rho_5$~=~100. Similar to 
Fig. \ref{fig:mu=0_cs2}, we have violation of causality
($c^2_s > 1$) and of the conformal limit ($c^2_s = 1/3$) at high temperatures.
In contrast, these limits 
are fully obeyed when using $\Lambda_T \to \infty$. 

\begin{figure}[h!] 
 \centering
 \includegraphics[width=0.45\textwidth]{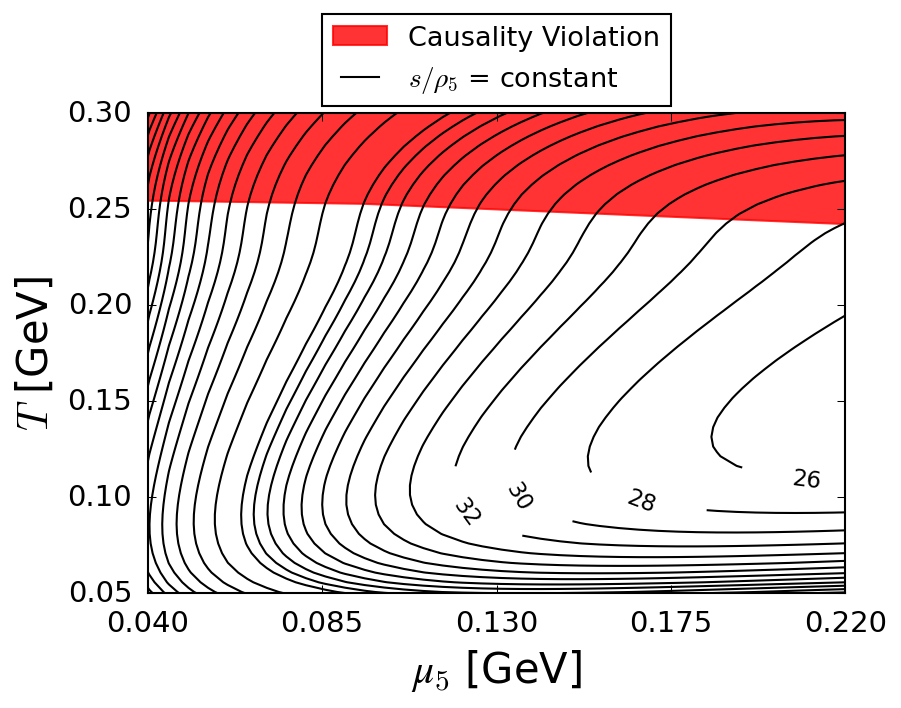}
 \caption{ Regions of causality violation in the $T - \mu_5$ plane with curves of constant entropy, $s/\rho_5$. }
 \label{fig:T_mu5_violation_curve}
\end{figure}

The red-shaded area in Fig.~\ref{fig:T_mu5_violation_curve}  indicates the region
in the $T - \mu_5$ plane, where $c^2_s > 1$  for different values of the $s/\rho_5$ ratio. The situation is similar to that 
in Fig.~\ref{fig:T_mu_violation_curve}, in that the temperature 
above which $c^2_s > 1$ decreases with the corresponding chemical potential.

\begin{figure}[h!] 
 \centering
 \includegraphics[width=0.45\textwidth]{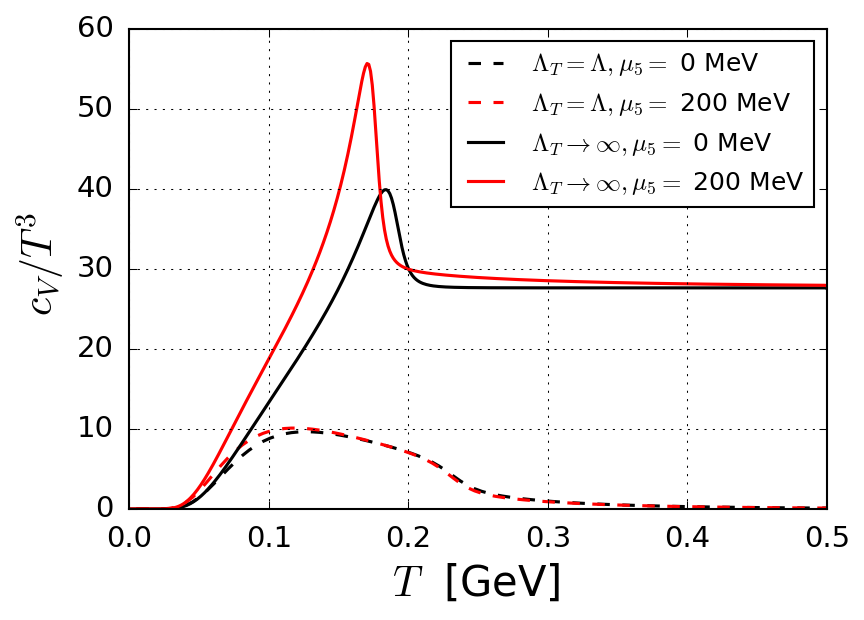}
 \caption{Specific heat divided by $T^3$ as a function of the temperature for two values of the chiral chemical potential $\mu_5$. }
 \label{fig:T_mu5_cv}
\end{figure}

\begin{figure}[h!] 
 \centering
 \includegraphics[width=0.45\textwidth]{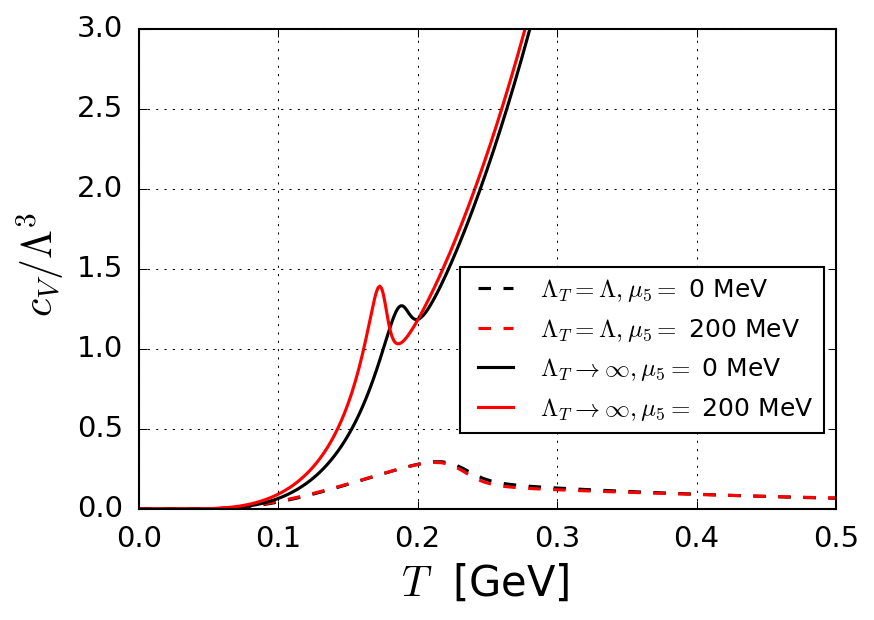}
 \caption{ Specific heat divided by $\Lambda^3$ as a function of the temperature for two values 
 of the chiral chemical potential $\mu_5$.}
 \label{fig:T_mu5_cv2}
\end{figure}

\begin{figure}[h!] 
 \centering
 \includegraphics[width=0.45\textwidth]{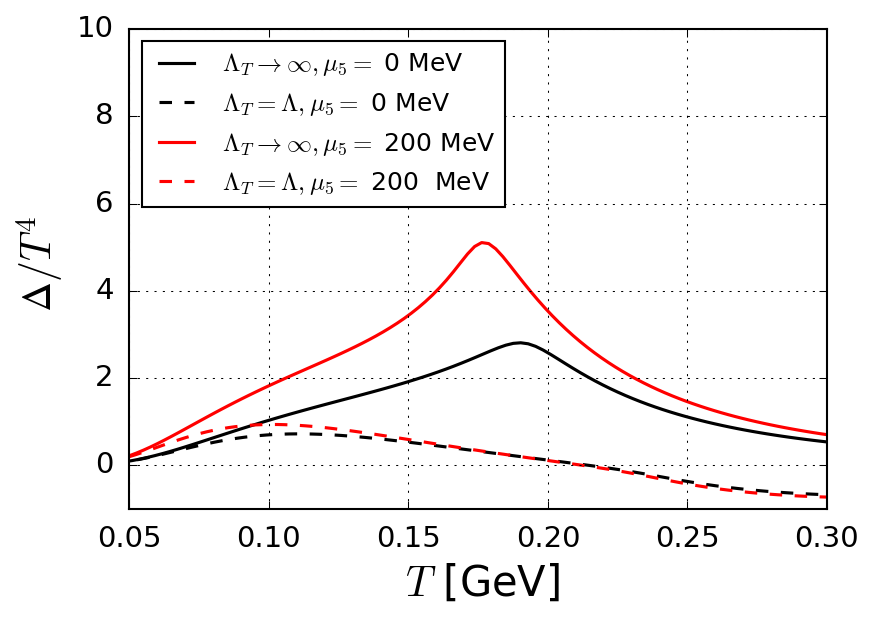}
 \caption{Results for normalized interaction measure (or trace anomaly) as a function of the temperature 
 for two values of the chiral chemical potential. }
 \label{fig:T_mu5_Delta}
\end{figure}

\begin{figure}[h!] 
 \centering
 \includegraphics[width=0.45\textwidth]{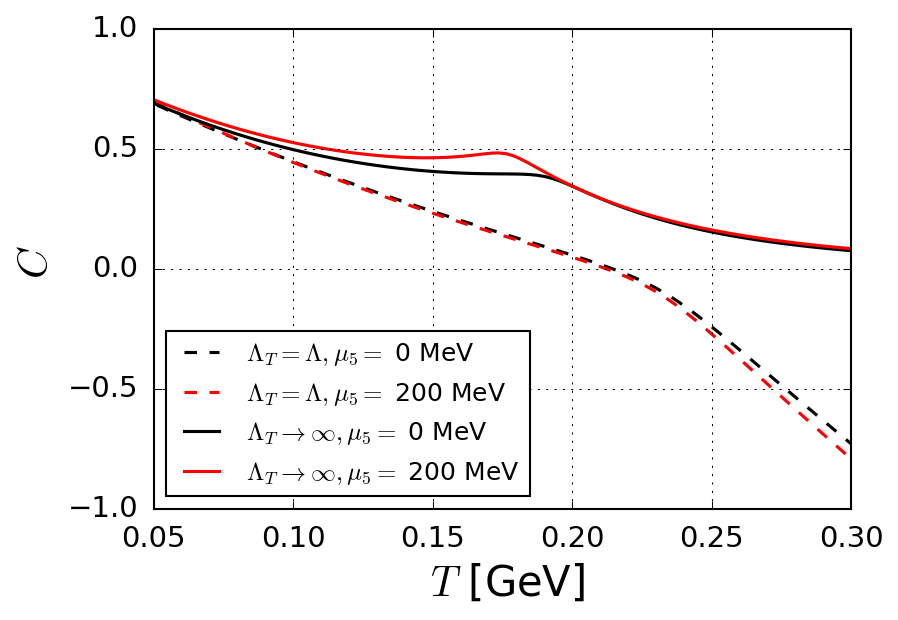}
 \caption{Conformal measure as a function of the temperature for two values of the
 chiral chemical potential. }
 \label{fig:T_mu5_Delta2}
\end{figure}

Next, we analyze the temperature dependence of the normalized specific heat
$C_V/T^3$ and $C_V/\Lambda^3$ regarding the use of the different cutoff strategies. 
Figures~\ref{fig:T_mu5_cv} and \ref{fig:T_mu5_cv2} display the results for $\mu_5=0$
and $\mu_5=200$~MeV. Again, the wrong high-temperature behavior 
of these quantities is clear when using $\Lambda_T = \Lambda$. The results also indicate a 
milder dependence of $C_V$ on $\mu_5$ than on $\mu_q$, using either $\Lambda_T = \Lambda$
or $\Lambda_T \rightarrow \infty$.

Finally, we plot the temperature dependence of $\Delta/T^4$ and $C=\Delta/\epsilon$,  
respectively, in Figs.~\ref{fig:T_mu5_Delta} and \ref{fig:T_mu5_Delta2}, for two values of
chiral chemical potential,  $\mu_5=0$ and $\mu_5=200$ MeV. These two quantities indicate very 
similar behavior when compared with Figs.~\ref{fig:T_mu_Delta} and \ref{fig:T_mu_Delta2}.
Clearly, the trace anomaly calculated with $\Lambda_T = \Lambda$  is negative for 
high temperatures. This does not happen when using $\Lambda_T \rightarrow \infty$. 
The conformal measure, with $\Lambda_T=\Lambda$, becomes negative at $T\sim200$ MeV, 
which is very close to the value at $\mu_5=0$. In this temperature region, $c^2_s > 1/3$. 
Since the $\mu_5=200$ MeV and $\mu_5=0$ situations are very similar, the conformal measure 
starts to disagree with the $\Lambda_T\rightarrow\infty$ case at $T\lesssim 100$ MeV, which indicates 
that the unphysical behavior induced by the regularization of the finite integrals  
happens for both $\mu_5=0$ and $\mu_5\neq 0$.

To conclude this section, we mention that regularization issues also 
affect the temperature and density-dependent quark condensate or, equivalently, the constituent
quark mass. For example, depending on the value of the cutoff, one can obtain negative
constituent quark masses at high values of the temperature. Therefore, to obtain physical values 
for the masses and condensate, the cutoff must be tuned accordingly.

\section{Conclusions}
In this work, we studied the implications of cutting moment integrals 
with a sharp three-momentum cutoff in the context of the two-flavor Nambu-Jona-Lasinio model. 
We compared the two procedures most commonly used in the literature: regularization
only of the vacuum divergent integrals or regularization of both divergent and finite
integrals with the same cutoff. For a systematic discussion concerning the importance 
of the choice of the regularization procedure, we calculated the physical observables for 
both zero quark and chiral chemical potentials, $\mu_q$=$\mu_5$=0, and finite quark 
and chiral chemical potentials. We explored scenarios relevant to heavy-ion collisions 
and compact stars. 

Our results indicated that causality violation and wrong high-temperature behavior of thermodynamical quantities are 
induced by the use of an ultraviolet cutoff in finite integrals depending on the 
Fermi-Dirac distributions. These problems happen in both $T-\mu$ and $T-\mu_5$ planes. 
On the other hand, when those finite integrals are not cut off, one obtains the expected
conformal limit in the speed of sound in the high-temperature limit. The specific heat, trace anomaly, 
and conformal measure were also investigated in these scenarios. In the specific heat and trace anomaly, 
the behavior of free quark gas is not achieved if one regulates integrals containing the thermal distributions, with 
the conformal measure presenting negative values for $T\gtrsim 250$  MeV.  

The analyses in this work indicate that, depending on the choice of the 
regularization, NJL-type models might induce artificial and unphysical behavior of physical 
observables. NJL-type models have been instrumental in guiding our intuition 
regarding chiral symmetry in vacuum and in matter; ignoring these regularization 
issues can lead to confusion and misguidance.     
%

\section*{ACKNOWLEDGMENTS}
This work was partially supported by Conselho Nacional de Desenvolvimento Cient\'ifico 
e Tecno\-l\'o\-gico  (CNPq), Grants Nos. 131212/2020-6 (A.E.B.P.), 309598/2020-6 (R.L.S.F.), 
304518/2019-0 (S.S.A.), 309262/2019-4; Coordena\c c\~{a}o  de Aperfei\c coamento de Pessoal de  
N\'{\i}vel Superior (CAPES) Finance  Code  001 and Fundação Carlos Chagas Filho de Amparo à Pesquisa do 
Estado do Rio de Janeiro (FAPERJ), Grant No.SEI-260003/019544/2022 (W.R.T); 
Funda\c{c}\~ao de Amparo \`a Pesquisa do Estado do Rio Grande do Sul (FAPERGS), Grants Nos. 19/2551- 0000690-0 
and 19/2551-0001948-3 (R.L.S.F.); Funda\c{c}\~ao de Amparo \`a Pesquisa do 
Estado de S\~ao Paulo (FAPESP), grant no. 2018/25225-9 (G.K.). The work is also part of the 
project Instituto Nacional de Ci\^encia e Tecnologia - F\'isica Nuclear e Aplica\c{c}\~oes 
(INCT - FNA), Grant No. 464898/2014-5. 

\bibliographystyle{spphys}
\bibliography{cs2.bib}

\end{document}